# Method of Self-Similar Load Balancing in Network Intrusion Detection System

Dmytro Ageyev, Lyudmyla Kirichenko, Tamara Radivilova, Maksym Tawalbeh
Kharkiv National University of Radioelectronics
Kharkiv, Ukraine
dmytro.aheiev@nure.ua, lyudmyla.kirichenko@nure.ua, tamara.radivilova@gmail.com, tavalbeh@icloud.com

Oleksii Baranovskyi
National Technical University of Ukraine Kiev Polytechnic Institute, Institute of physics and techlonogy
Kyiv, Ukraine
alexey.baranovskiy@gmail.com

*Abstract*— In this paper, the problem of load balancing in network intrusion detection system is considered. Load balancing method based on work of several components of network intrusion detection system and on the analysis of multifractal properties of incoming traffic is proposed. The proposed method takes into account a degree of multifractality for calculation of deep packet inspection time, on the basis of which the time necessary for comparing the packet with the signatures is calculated. Load balancing rules are generated using the estimated average deep packet inspection time and the multifractality parameters of incoming load. Comparative analysis of the proposed load balancing method with the standard one showed that the proposed method improves the quality of service parameters and the percentage of packets that are not analyzed.

## I. Introduction

Intrusion detection (attack) is the process of monitoring events occurring in a computer system or network in order to search for signs of possible incidents. Intrusion detection system (IDS/IPS, Intrusion detection system/Intrusion prevention system) is a necessary element of protection against network attacks. The main purpose of such systems is to identify the facts of unauthorized access to corporate network and take appropriate countermeasures [1,2].

Currently, there are works aimed at solving the problem of load balancing in Network Intrusion Detection System (NIDS) [3-5]. In [6] the parallel architecture of NIDS is considered, which overcomes the restriction on intrusion detection, distributing network traffic load on an array of sensor nodes. The paper [7] proposes a general architecture for deploying NIDS in a network that uses three scaling options: path-by-division for sharing responsibilities, replication of traffic to NIDS clusters, and aggregation of intermediate results to separate costly NIDS processing. In [3] authors offer various policies for activating/deactivating a dynamic load balancer by comparing single and double threshold circuits and load representations based on resource models and load aggregation models.

The aim of this work is a modification of the load balancing method taking into account the self-similar properties of the input load in NIDS.

## II. Description of NIDS

Since an architecture based on only one traffic sensor can not be sufficient to resist vulnerabilities in networks that are characterized by a large amount of traffic, therefore, a distributed architecture with multiple sensors is the most efficient solution for analyzing traffic and high-speed networks [8-10]. This distributed architecture is characterized by a set of nodes that direct portions of network traffic to different NIDS sensors through some traffic shaping policy [9, 11]. Each NIDS sensor analyzes the received traffic for intrusions.

The main problem is that the incoming traffic received by a distributed architecture of NIDS has a long-term dependency and bursts. Therefore, the work uses analysis of incoming traffic for the presence of fractal properties and dynamic redistribution of the load among a sensors [12, 13]. It is proposed to use a balancer that receives periodic information about sensors condition, and based on some policy, it can implement a load balancing mechanism to move part of network traffic from overloaded sensors to less loaded [12,13]. Load conditions of each sensor are usually evaluated by analyzing incoming traffic. However, in conditions of intense traffic with unexpected bursts, it is extremely difficult to determine the optimal decision-making policy and the load-balancing algorithm for the load balancing process.

At each time point $t \in [t_0, t_0 + T]$ the input of the NIDS receives traffic with intensity $\lambda = [\lambda_1, \lambda_2, ..., \lambda_\sigma]$, belonging to the $qs$-th class of service, which must be delivered to $i$-th node $Nids_i$ for processing or any further transfer, not exceeding the specified maximum permissible delay values and percentage of losses, depending on the throughput and the current load of nodes at a particular time.

Traffic has many characteristics $V = \{\lambda, h, S_t\}$, where $\lambda = [\lambda_1, \lambda_2, ..., \lambda_\sigma]$ is the intensity of incoming flows (packets); $h = [H, h(q), \Delta h]$, where $h(q)$ is the sample value of the function of the generalized Hurst exponent, $H = h(2)$ is the value of the Hurst parameter, $\Delta h = h(q_{min}) - h(q_{max})$ is the range of values of the generalized Hurst index for the traffic section; the set $S_t = \{st_1, ..., st_h\}$ is collected information about NIDS network traffic, where $st_t$ can match network address, port, network packet field values, device hardware address, protocol identifiers, packet field size, etc.

To describe the set of used signatures, we introduce the set $Sg = \{Sg_1, Sg_2, ..., Sg_n\}$, where $Sg_j$ are the elements of set of signatures from database of NIDS signatures. Then a set of rules for responding to NIDS on alleged invasions $R = \{R_1, R_2, ..., R_n\}$, where $R_j$ is a rule of allowing/prohibiting response to a specific type of intrusion can be divided into two parts.

Rules that allow to pass packets of $R^+ = \{R_1, R_2, ..., R_k\}$ type, corresponding to signatures from a subset $Sg^+ = \{Sg_1, Sg_2, ..., Sg_k\}$ of a set $Sg$, and rules that prohibit to pass packets of $R^- = \{R_1, R_2, ..., R_s\}$ type corresponding to signatures from a subset $Sg^- = \{Sg_1, Sg_2, ..., Sg_s\}$ of a set $Sg$, $R^+ \cap R^- = \varnothing$ and $Sg^+ \cap Sg^- = \varnothing$.

An input node $Nids_i$ receives multiple independent multifractal packet flows with different intensity $\lambda_1, \lambda_2, ..., \lambda_\sigma$, which are distributed between nodes in accordance with a policy of traffic shaping in limited capacity queues $Q_i$ [8,9]. The NIDS system contains of $N$ ($i = \overline{1, N}$) nodes, each of them has its own queue, as well as one input queue.

To determine load state of NIDS node it is necessary to collect statistics of input queue for a period of time $[t_0, t_0 + T]$.

The average CPU load of $i$-th NIDS node is $CPU_i^{n_i}$. Similarly, average rate of memory utilization is $RAM_i^{m_i}$, available bandwidth of channel of $i$-th NIDS node is $Net_i^{k_i}$.

The total imbalance values of all NIDS nodes are defined as: $IMB_{tot} = \frac{1}{N}\sum_i^N IMB_i$, where imbalance values of the $i$-th NIDS nodes:
$$IMB_i = K_c(CPU_i^u - CPU_u^{All})^2 + K_r(RAM_i^r - RAM_r^{All})^2 + K_n(Net_i^k - Net_k^{All})^2,$$
$CPU_u^{All}$ is the average rate of all processors utilization in system, $RAM_r^{All}$ is average rate of all memory and average available throughput of channel in the system $Net_k^{All}$, $n_i$ are the numbers of CPUs on $i$-th node, the parameters $K_c, K_r, K_n$ denote the weighting coefficients for processor, memory, and available network bandwidth respectively, that are selected experimentally in such a way that $K_c + K_r + K_n = 1$ and depending on the performed tasks and system structure.

Packet service time $T_{serv}$ refers to the operation time, during which a comparison is made between the signatures $Sg = \{Sg_1, Sg_2, ..., Sg_n\}$ and one or more packets for intrusion detection $THR = \{THR_1, THR_2, ..., THR_k\}$. Each node compares each or several packets with one or more signatures $Sg_j$. The more signatures are present for comparison, the more time $T_{serv}$ is needed for analysis. In other words, the packet service time is proportional to the number of signatures that must be matched with a packet. The ratio of the packet service time in one node to the total packet service time for an operation on all nodes is called the packet transfer rate.

Average service time of IDS is time, required by IDS with configuration of a rules $R_j$ to successfully determine a permission/prohibition of responding to a specific type of intrusion. The average IDS service time is determined as [17]:
$$T_{serv}^{IDS} = \left[\sum_{j=1}^N P(j)\sum_{s=1}^N T_{serv}(s)R^-(s)\right] + \left(1 - \sum_{j=1}^N P(j)\right) \times \sum_{j=1}^N T_{serv}(j)R^+(j),$$
where $P(j)$ is probability of blocking rules $R_j$; each rule $R_s$ is responsible for only one type of malicious event.

A service time-aware load balancing method uses a deep packet inspection (DPI) time of the NIDS node in which a connection between traffic and signature is analyzed using the estimated packet service time. The DPI time for one packet is determined by the number of signatures corresponding to packet.

III. PROPOSED METHOD

In this paper we propose a load balancing method based on service time-aware load balancing, and signature analysis, and taking into account the properties of self-similar traffic. Self-similarity of traffic means maintaining the distribution law for different time scales [10,14]. The degree of self-similarity is characterized by a number - Hurst exponent H (0<H<1), which also is a measure of long-term dependence. The greater the value of H, the stronger and longer correlation between traffic values; this property of traffic does not allow a system resources to be quickly released, because a large data values are followed by large ones [12, 13, 15].

Burst property (heterogeneity of traffic) is characterized by the presence of large emissions in a traffic realization with a small intensity. For the mathematical description of burst property, it is necessary to consider the $q$-th moments ($q>2$) of a process. The characteristic of traffic multifractal properties is the generalized Hurst exponent $h(q)$. It is nonlinear function, that is based on the $q$-th moments and characterizes a heterogeneity of self-similar traffic. The larger the range $\triangle h = h(q_{min}) - h(q_{max})$, the greater the heterogeneity (bursts) of traffic, i.e. a stronger emissions are present in a traffic realization. The method proposed in this paper can provide uniform load balancing at discrete instants in order to fully utilize the multithreaded NIDS, which leads to more efficient use of the system in processing data for intrusion detection. This method consists of the following operations.

1. Customers receive a lot of data packets of intensity $\lambda = [\lambda_1, \lambda_2, ..., \lambda_\sigma]$, which belong to the $qs$-th class of service. The input data has many characteristics $V = \{\lambda, h, S_t\}$ and necessarily contain a type of protocol (Transmission Control Protocol – TCP, User Datagram Protocol – UDP, etc.) and protocol property. The properties of data packet protocol contain the source IP address, the source port, the destination IP address and the destination port.

2. The arrived packets are balanced on the node according to their arrival rate $T_{serv}/T_{serv}^{IDS}$ IDS for a specified period of time $T$. Balancing can be applied for any chosen algorithm. The incoming data packets are processed by a partitioning procedure that classifies them into data type flows of a type of service by identifying the headers of arrived packets in accordance with a type and protocol property.

3. For each flow there are calculated a multifractal parameters $\Delta h = h(q_{min}) - h(q_{max})$, $H$ and the ratio of signatures number or every qs-th flow type to the total signatures number $Sg = \{Sg_1, Sg_2, ..., Sg_n\}$.

4. The balancer analyzes incoming packets at predetermined time periods T. The ratio of the packets number for each type of service qs-th to the total number of arrived packets for a specified time period is calculated. Next, the balancer evaluates a packet comparison time with a signature $T_{serv}$ and a processing time for a specific service, based on a multifractality parameters, and the ratio of the number of signatures for each type of flow to the total number of signatures $Sg = \{Sg_1, Sg_2, ..., Sg_n\}$.

5. An intrusion detection procedure is performed in accordance with a signature set $Sg = \{Sg_1, Sg_2, ..., Sg_n\}$. That is, a load balancer compares at least one of the signatures with a load of arrived packets at least one type of service.

6. The average time of DPI packets $T_{new}^{qs}(qs)$ corresponding to a particular service is evaluated. The average time of DPI packets for signatures of a particular service can be calculated using a signatures number and the average processing time of all signatures of a particular service, taking into account the multifractality parameters.

$$T_{new}^{qs} = \begin{cases} T_{serv}^{IDS}(qs), & H = 0,5; \\ T_{serv}^{IDS}(qs) + (H - 0.5)T_{serv}, & 0.5 < H < 0.9, \Delta h \leq 0.4; \\ T_{serv}^{IDS}(qs) + (H - 0.5)(\Delta h - 0.4)T_{serv}, & 0.5 < H < 0.9, 0.4 < \Delta h < 1; \\ T_{serv}^{IDS}(qs) + T_{serv}, & H \geq 0.9 \text{ or } H > 0.5, \Delta h \geq 1, \end{cases}$$

where $T_{serv}^{IDS}(qs)$ is determined according to a class of service and a necessary resources, the value $T_{serv}$ is calculated depending on a signatures number. Estimating an average time of DPI packets does not change ($T_{new}^{qs} = T_{serv}^{IDS}(qs)$) if traffic is a normal Poisson flow ($H = 0.5$). With a value $0.5 < H < 0.9$ and a low data dispersion ($\Delta h \leq 0.4$), the value $T_{new}^{qs}(qs)$ increases in proportion to the value of the Hurst index. When the value of the Hurst parameter $0.5 < H < 0.9$ and a high data dispersion ($0.4 < \Delta h < 1$) values $T_{new}^{qs}(qs)$ increase in proportion to both parameters. Estimating the average time of DPI packets with the maximum value $T_{serv}^{IDS}(qs) + T_{serv}$ is obtained for the value $H \geq 0.9$ or for persistent traffic ($H > 0.5$) with a range of values of the generalized Hurst exponent $\Delta h \geq 1$. Also, the average of all signatures processing time can be reflected as weight when estimating the average DPI time.

7. Generating a first list of service, which has an average DPI time greater than or equal to a specified level, by sorting. And, accordingly, generate a second list of service, which has an average DPI time less than a specified level.

8. Changing the balancing rule is initiated to turn-on. Changing occurs, when a specified condition that affects a total DPI time of a distributed NIDS is met, or the specified time for updating a signature database expires.

9. The runtime statistics that are needed to compare a signatures with a package for each type of service are recorded.

10. A new load balancing rule $Load_T(T_{new}^{qs}, H, \Delta h)$ is created based on an estimate of a mean DPI $T_{serv}^{IDS}(qs)$. It can change periodically depending on result of traffic analysis. According to a new load balancing rule, packets for each type of service included in the first list of service are assigned for processing to certain NIDS components, and packets for service types included in the second list of service are assigned for processing to other NIDS components.

11. After the signature analysis, a load balancing rule is updated. Load balancing is performed in accordance with a updated load balancing rule $Load_T$.

12. Balancing of arrived packets are carrying out in next set time period $2T$ on several NIDS components using newly created balancing rule based on a result of analysis of packets arriving at a specified time period $T$.

13. Events of exceeding the specified level by the traffic of incoming packets or the end of specified load balancing time are monitored before the analysis of packets arriving in the $2T$ time period. Packets are processed and analyzed by components NIDS using the updated load balancing rule.

14. Follow balancing in accordance with operations 2-13.

Thus, the method of load balancing in network IDS, that takes into account the self-similar properties of incoming traffic is presented in this paper.

## IV. SIMULATION RESULTS

Simulation was performed in a program written in Python to test the validity of proposed method. During simulation, we assigned an equal processing time for all rules (one time unit). The input of the system was fed with the generated multifractal traffic described in [4]. Data coming from an external network creates additive multifractal traffic that contains threat markers. These data are sent to a balancer, which regulates the data flow using selected balancing policy and are sent to the NIDS nodes.

To analyze the proposed balancing method, numerous studies of NIDS balancing system were performed for different values of multifractal traffic parameters: the range of values of the generalized Hurst exponent $1,5 \leq \Delta h \leq 6$, the value of the Hurst parameter $0,6 \leq H \leq 0,9$, and the intensity of application flow $0,5 \leq \lambda \leq 1$. Table 1 shows the values of performance indicators for the standard load balancing method (SM) and the proposed method (PM) for different values of the multifractality parameters of input traffic.

TABLE I. CHANGE OF VALUES OF PERFORMANCE INDICATORS DEPENDING ON MULTIFRACTALITY PARAMETERS

| Traffic parametrs | Packet loss | | IMB | | Not analyzed packets | |
|---|---|---|---|---|---|---|
| | SM | PM | SM | PM | SM | PM |
| H=0,6, Δh=2 | 3.4 | 1.9 | 0.28 | 0.21 | 8.9 | 7.8 |
| H=0,6, Δh=6 | 6.6 | 6.3 | 0.66 | 0.57 | 20 | 17.2 |
| H=0,7, Δh=2 | 4.6 | 4 | 0.42 | 0.34 | 12.4 | 10.3 |
| H=0,7, Δh=6 | 11 | 9.8 | 0.7 | 0.62 | 24.5 | 22 |
| H=0,8, Δh=2 | 7.6 | 7.1 | 0.51 | 0.44 | 16 | 16 |
| H=0,8, Δh=6 | 16.6 | 15.9 | 0.81 | 0.72 | 32.4 | 28 |
| H=0,9, Δh=2 | 12.1 | 11.2 | 0.5 | 0.46 | 19 | 19 |
| H=0,9, Δh=6 | 23.1 | 22 | 0.99 | 0.92 | 39.7 | 35.1 |

Studies have shown that multifractal traffic characteristics significantly affect the system imbalance, the number of not analyzed packages and lost data increases, as it is shown in Table 1. As small values of H and small heterogeneity a balancing system comes in equilibrium state and the performance of NIDS is satisfying, and imbalance value tends to zero. As large values of the Hurst index and large heterogeneity, the balancing system is in an unstable state and the imbalance value changes several times, which leads to the maximum load of resources, and consequently to a significant increase in the number of not analyzed packets and lost data.

## CONCLUSION

In this paper, we propose a new approach to solving the problem of self-similar load balancing in high-speed Intrusion Detection Systems. The paper proposes a modified load balancing method, based on accounting of service time, in which packets arriving in a specified time period are compared with one or more signatures. The proposed method takes into account the degree of traffic multifractality for the calculation of DPI time, on the basis of which the time required to compare a packet with signatures is computed, collects work time statistics, generates and updates the rules for balancing incoming packets. The proposed load balancing method provides a statically uniform load distribution on NIDS nodes, a low percentage of lost data and not analyzed packets and aims to provide high rate and accuracy of intrusion detection with a quality balancing of the incoming load.

In the future work we planned to carry out a comparative analysis of work results of rules for intrusion detection by signatures and abnormal behavior for various types of attacks (denial of service, suspicious activity, system attack), their influence on the multifractal characteristics of traffic and determining boundary values for behavior characteristics of the input data to reduce the probability of the error detection.

REFERENCES

[1] B. Mukherjee, L. T. Heberlein and K. N. Levitt, "Network intrusion detection," in IEEE Network, vol. 8, no. 3, pp. 26-41, May-June 1994.

[2] C. Warrender, S. Forrest and B. Pearlmutter, "Detecting intrusions using system calls: alternative data models," Proceedings of the 1999 IEEE Symposium on Security and Privacy (Cat. No.99CB36344), Oakland, CA, 1999, pp. 133-145.

[3] M. Andreolini, S. Casolari, M. Colajanni and M. Marchetti, "Dynamic load balancing for network intrusion detection systems based on distributed architectures," Sixth IEEE International Symposium on Network Computing and Applications (NCA 2007), Cambridge, MA, 2007, pp. 153-160. doi: 10.1109/NCA.2007.17

[4] I. Ivanisenko, L. Kirichenko and T. Radivilova, "Investigation of self-similar properties of additive data traffic," 2015 Xth International Scientific and Technical Conference "Computer Sciences and Information Technologies" (CSIT), Lviv, 2015, pp. 169-171.

[5] D. Ageyev and N. Qasim, "LTE EPS network with self-similar traffic modeling for performance analysis," 2015 Second International Scientific-Practical Conference Problems of Infocommunications Science and Technology (PIC S&T), Kharkiv, 2015, pp. 275-277.

[6] D. Ageyev and M. Salah, "Parametric Synthesis of Overlay Networks with Self-Similar Traffic", Telecommunications and Radio Engineering, vol. 75, no. 14, pp. 1231-1241, 2016.

[7] Lambert Schaelicke, Kyle Wheeler and Curt Freeland. "SPANIDS: A Scalable Network Intrusion Detection Loadbalancer", Proceeding CF '05 Proceedings of the 2nd conference on Computing frontiers, pp. 315-322, 2005.

[8] Victor Heorhiadi, Michael K. Reiter and Vyas Sekar. "New Opportunities for Load Balancing in Network-Wide Intrusion Detection Systems", Proceeding CoNEXT '12 Proceedings of the 8th international conference on Emerging networking experiments and technologies, pp. 361-372, 2012. Doi: 10.1145/2413176.2413218

[9] Xiao-Qian Li and Tom Chen. "Load balancing method for Network Intrusion Detection", United States Patent Application Publication, US 2010/0246592 A1, Sep. 30, 2010.

[10] Yoon-ho Choi, Seung-Woo Seo, Bon-Hyun Koo and Hye-Jung Cho. "Load balancing method and apparatus in Intrusion Detection Sysytem", United States Patent Application Publication, US 2017/0295191 A1, Oct. 12, 2017.

[11] A. Erramilli, M. Roughan, D. Veitch and W. Willinger, "Self-similar traffic and network dynamics," in Proceedings of the IEEE, vol. 90, no. 5, pp. 800-819, May 2002. doi: 10.1109/JPROC.2002.1015008

[12] O. Lemeshko and O. Yeremenko. "Enhanced method of fast re-routing with load balancing in software-defined networks." Journal of Electrical Engineering, vol. 68, iss. 6, pp. 444-454, December 2017.

[13] L. Kirichenko and T. Radivilova, "Analyzes of the distributed system load with multifractal input data flows," 2017 14th International Conference The Experience of Designing and Application of CAD Systems in Microelectronics (CADSM), Lviv, 2017, pp. 260-264.

[14] I. Ivanisenko and T. Radivilova, "The multifractal load balancing method," 2015 Second International Scientific-Practical Conference Problems of Infocommunications Science and Technology (PIC S&T), Kharkiv, 2015, pp. 122-123.

[15] A. Yerokhin, O. Turuta, A. Babii, A. Nechyporenko and I. Mahdalina, "Usage of phase space diagram to finding significant features of rhinomanometric signals," 2016 XIth International Scientific and Technical Conference Computer Sciences and Information Technologies (CSIT), Lviv, 2016, pp. 70-72.

[16] L. Kirichenko, I. Ivanisenko and T. Radivilova, "Dynamic load balancing algorithm of distributed systems," 2016 13th International Conference on Modern Problems of Radio Engineering, Telecommunications and Computer Science (TCSET), Lviv, 2016, pp. 515-518.

[17] K. Alsubhi, N. Bouabdallah and R. Boutaba, "Performance analysis in Intrusion Detection and Prevention Systems," 12th IFIP/IEEE International Symposium on Integrated Network Management (IM 2011) and Workshops, Dublin, 2011, pp. 369-376.

[18] Yu Ying and Deng Qidong. A Dynamic Forecast Load-balancing Algorithm for High-speed. Network Instruction Detection System.